# LOW TEMPERATURE INVESTIGATION OF ELECTRICAL CONDUCTION IN POLYSILICON: SIMULATION AND EXPERIMENT

*S. Ecoffey (1), S. Mahapatra (1), V. Pott (1), D. Bouvet (1), G. Reimbold (2) and A. M. Ionescu (1)*

(1) Ecole Polytechnique Fédérale de Lausanne (EPFL), STI – IMM – LEG, 1015 Lausanne, Suisse.
(2) CEA-LETI, 17 rue des Martyrs, 38054 Grenoble Cedex 9, France.

## ABSTRACT

Investigation of electrical conduction in polysilicon nanowires (polySiNW) with nanograins (5 to 20nm), based on Monte Carlo (MC) simulations and electrical measurements from 4K to 300K are presented. Some irregular Coulomb Oscillations (CO) are observed at temperatures lower than 200K showing several period widths due to the random distribution in grain size (5-20nm). A remarkable result consists in more effective oscillations observed at intermediate range of temperatures (between 25K and 150K) and high drain voltages. The temperature dependence of COs is explained by the fact that in a multiple asymmetric dot system at low temperature, COs are observed not at the lowest but at an intermediate temperature range, whereas the drain voltage dependence is due to an enhanced non-resonant tunneling. MC simulations have confirmed experimental observations.

## 1. INTRODUCTION

The trend of Moore's law has been followed for more than thirty years and the ITRS roadmap predicts that it will last for approximately a decade [1]. However, this constant and aggressive scaling of CMOS will reach basic physical limits: the size of atoms. New solutions in terms of device architectures [2], and process materials [3] must be found if one wants to go beyond CMOS. In that way, many different emerging technologies have been proposed to replace and/or complement CMOS in the future: molecular electronic, plastic electronic, Single Electron Transistor (SET), Quantum Cellular Automata, Nano-Electro-Mechanical-Systems, etc. Other promising candidates are the nanowires that have attracted high interest due to their potential to realize new building blocks for nanoelectronics and their ability to test the electrical conduction at the nanoscale [4,5]. Moreover, silicon nanowires benefit for a huge and well established experience on silicon and present interesting Coulomb Blockade (CB) [6,7] or SET effects [7] for both logic and memory applications.

This works reports on the electrical behavior of ultra-thin (10nm) gated polySiNWs with grain sizes ranging from 5 to 20nm. MC simulations and experimental measurements are coherent and a first order theory is proposed.

## 2. MC SIMULATIONS OF POLYSINW AT SUB-AMBIENT TEMPERATURES

Fig. 1 depicts the gated-polySiNW device fabricated for the electrical characterization of ultra-thin nanograin polySi wires presented in the next chapter. It operates as a pseudo-MOS structure with n-type buried gate and polySiNW channel. The substrate is <100> p-type Si. Two thick (100nm) polysilicon pads have been structured in the contact regions in order to preserve the 20nm gate oxide when contacting. The channel of the transistor is made out of deepUV structured ultra-thin polysilicon film with grain sizes ranging from 5 to 20nm and equivalent phosphorous doping of ~$3x10^{19}$. A more detailed description of the technological process and the polysilicon physical properties are given in [8] and [9] respectively.

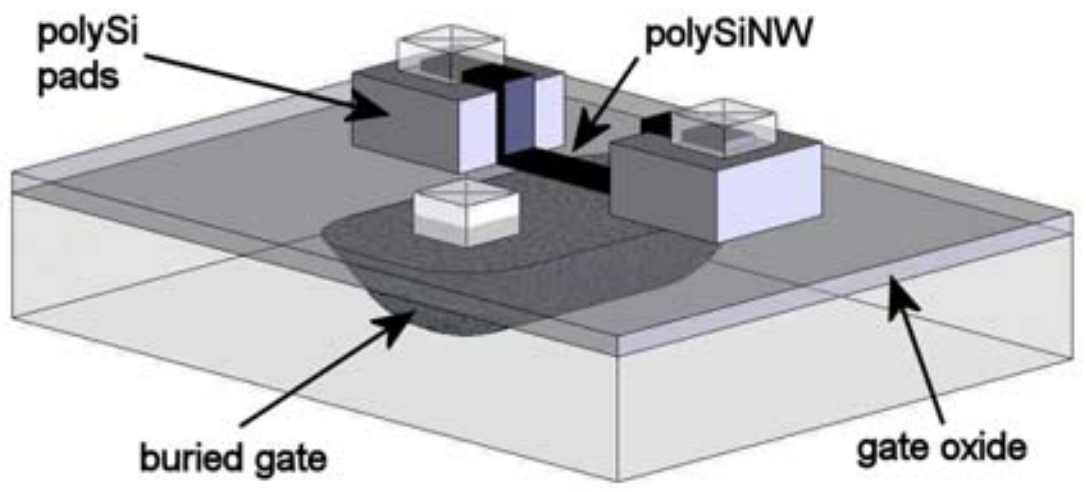

***Fig. 1:*** *3D sketch of the gated polySiNW device.*

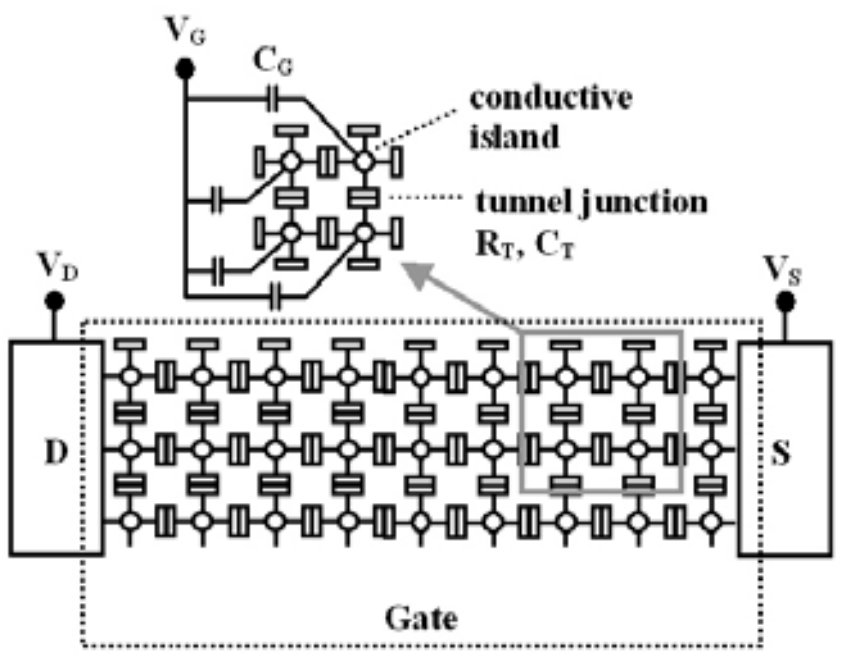

***Fig. 2:*** *electrical equivalent schematic of a gated polySiNW considering a (3x9) array with zoom on a (2x2)array of islands to illustrate the gate bias applied on each grain.*

Simulations have been carried out using SIMON, a MC simulator [10]. The intrinsic structure of a lithographically defined polySiNW can be approximated by a large array of connected Si dots separated by tunnel junctions. The equivalent device of a gated polySiNW used for simulations is presented in Fig. 2. The doped polysilicon grains are modelled as conductive islands connected to each other by tunnel junctions and share a common gate. For the simulations, tunnel junction capacitances, $C_T$, and gate capacitances, $C_G$, are taken to be 1aF for 20nm grains and 0.5aF for 10nm grains, tunnel junction resistances are 1MΩ. These values are used only for first order estimations and have been derived from rough geometrical assumptions.

Fig. 3 shows $I_D$-$V_{GS}$ characteristic of a 5x15 dot array structure (equivalent grain size of 20nm) at different temperatures, ranging from 1K up to 100K. It demonstrates that, such architecture provides clear CO, the number of oscillations increasing with temperature. This type of stochastic CO has already been observed as being specific to double [11], and multiple dot systems [12]. In a multiple dot system, at low temperature, COs are observed not at the lowest but at an intermediate temperature range. The electrochemical potential of each dot must be equivalent, within the limits forced by thermal smearing:

$$\left| eV_G-(n_m+1/2)\,\Delta m \right| \leq k_BT \qquad n_m=1, 2, \ldots \qquad (1)$$

where $V_G$ is the gate voltage, $n_m$ is the number of electrons on the m$^{th}$ dot, $\Delta m=e^2/C_{gm}$, and $k_BT$ is the thermal energy. Thus, with increasing temperature, the number of $V_G$ values where the condition is fulfilled is also increasing.

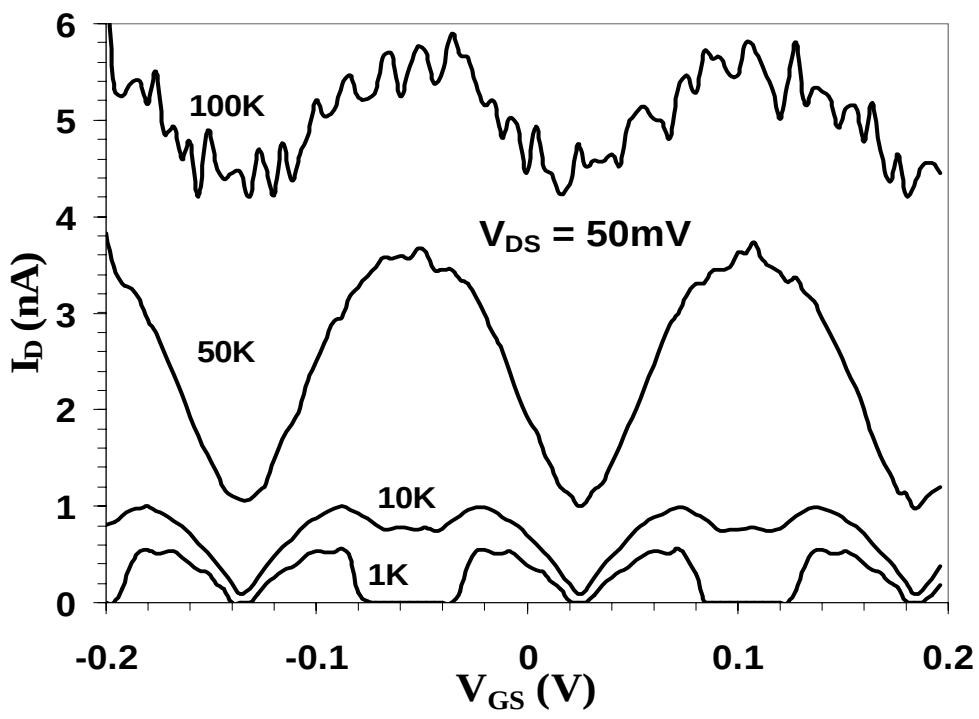

***Fig. 3:*** *MC simulated CO at different temperatures of a (5x15) array of 20nm polySi grains ($C_G$=$C_T$=1aF).*

The choice of the adequate oscillating conditions are even more difficult because optimal drain voltage conditions are needed. Wang *et al* [13], have observed that the number and periodicity of stochastic COs in an asymmetric double dot system are drain voltage dependent due to the enhanced non-resonant tunneling. This effect has been confirmed by Kawamura *et al* [12] for a multiple asymmetric dots system: a polycrystalline silicon film. Fig. 4 illustrates the dependence of the periodicity and number of COs over the drain voltage of a multiple dot system: (5x15) array of 20nm equivalent grains. This effect is less apparent compared to the example given before [12,13] because the system is symmetric. It will be more obviously shown with the electrical measurements of the polySiNWs.

A much realistic case corresponding to our gated polySiNW is a random mixture of grain size of 10 and 20nm. Fig. 5 compares simulated $I_D$-$V_{GS}$ characteristic of three type of polySiNW at 10K: (i) a 5x15 array with 20nm dots ($C_G$=$C_T$=1aF), (ii) a 5x15 array with 10nm dots ($C_G$=$C_T$=0.5aF), and (iii) a 5x15 array random mixing 20nm ($C_G$=$C_T$=1aF) and 10nm ($C_G$=1aF≠$C_T$=0.5aF) dots.

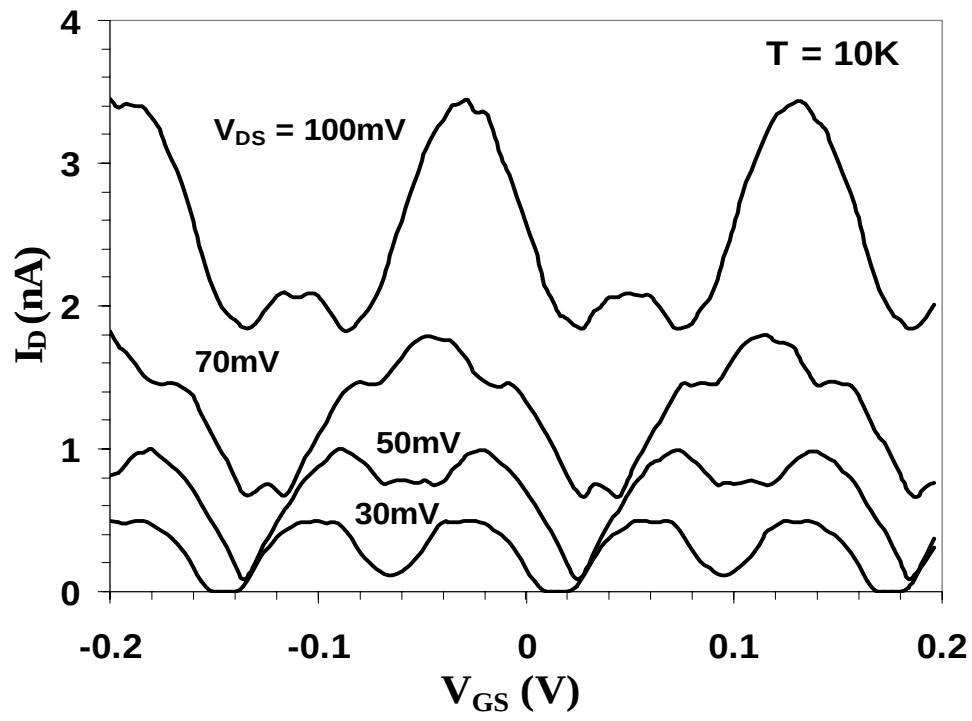

***Fig. 4:*** *MC simulated CO at different drain voltages of a (5x15) array of 20nm polySi grains ($C_G$=$C_T$=1aF).*

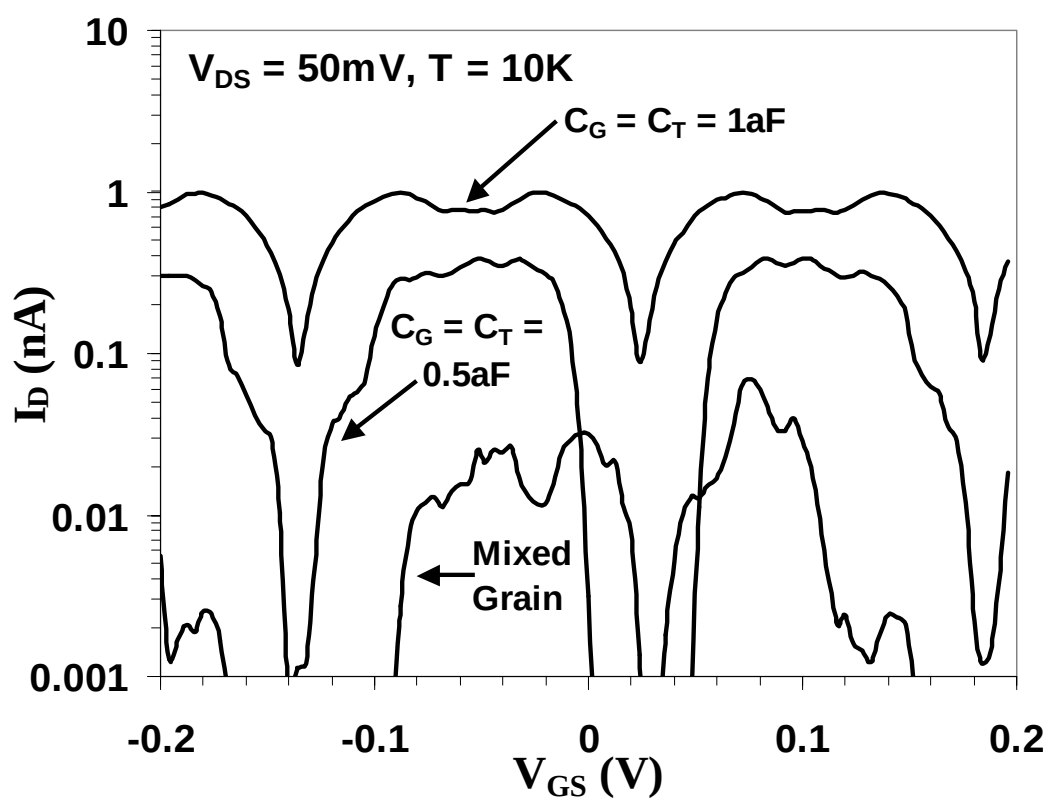

***Fig. 5:*** *MC simulated CO characteristics of a (5x15) array of: (i) 20nm ($C_G$=$C_T$=1aF), (ii) 10nm ($C_G$=$C_T$=0.5aF) and (iii) randomly mixed grains.*

It should be noted that due to the simulator used (free online version of SIMON) the gate capacitances $C_G$ of the random mixing of grains have been set to 1aF for each grain. Fig. 4 shows when using the same biasing conditions and temperature, the periodicity measured for polySiNWs with grains of the same size disappears in a wire with variable (randomly mixed) grain sizes.

## 3. ELECTRICAL MEASUREMENTS

The $ID_{NW}$-$VD_{NW}$ characteristics at constant gate voltage, down to a temperature of 4.1K is depicted on Fig. 6. The non-ohmic plateau is probably due to the superposition of two phenomena: (i) probable Schottky nature of the contacts, and (ii) evident Coulomb Blockade in the polySiNW. As the temperature is decreased from 75K down to 4.1K the width of the plateau is widening due to a more effective Coulomb Blockade in the wire.

Fig. 7 represents an $ID_{NW}$-$VG_{NW}$ characteristic at 25K and 4.1K. The solid black line shows the tendency curve after a noise reduction operation consisting of averaging each measured points with the two previous and two next measured points. It should also be mentioned that this characteristic has been carried out with a measurement step of 1mV. The two curves show clear oscillations, however, no systematic period could be extracted from those points. In fact, the choice of the adequate temperature for a large number of $VG_{NW}$ is difficult due to the broad dispersion of grain sizes between 5 and 20nm, and the relatively wide dimensions of the wires (compared to the grains).

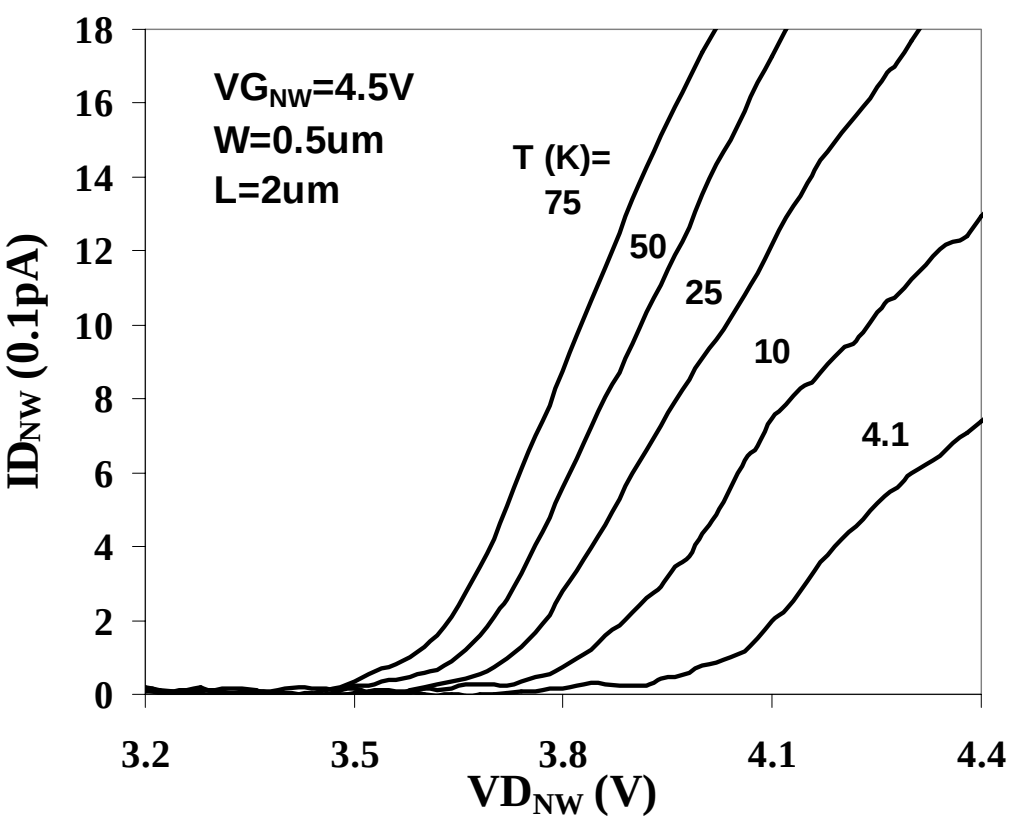

***Fig. 6:*** *measured $ID_{NW}$-$VD_{NW}$ of a gated polySiNW showing a temperature dependent Coulomb Gap.*

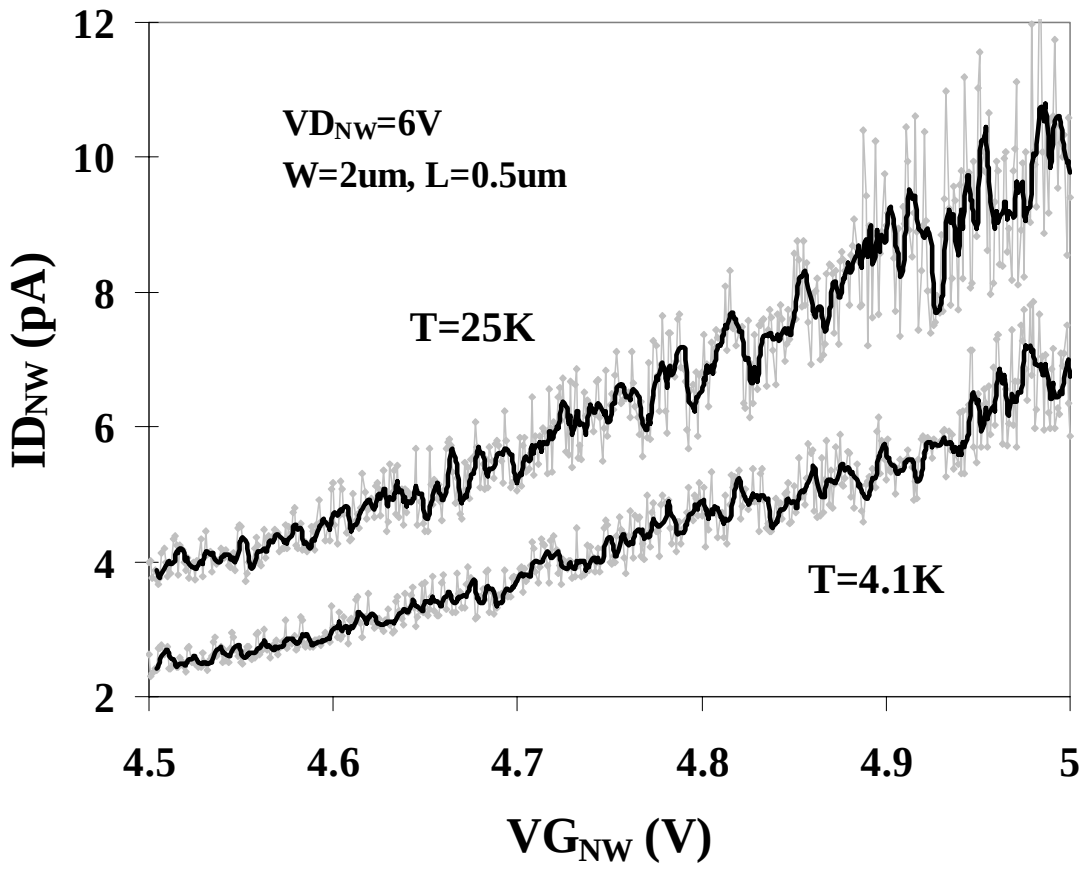

***Fig. 7:*** *measured $ID_{NW}$-$VG_{NW}$ characteristic of a polySiNW at 25K and 4.1K (meas. step 1mV).*

Nevertheless, Fig. 7 shows that there is an optimal temperature for which the number and size of oscillations are maximum (maximum COs are not obtained at lowest temperature), as previously shown by MC simulations and predicted by Kawamura et al [12]. It tends to demonstrate that these oscillations are really stochastic COs.

The stochastic COs are confirmed when looking at the characteristics on Fig. 8 where $ID_{NW}$-$VG_{NW}$ are represented for four different drain voltages. The plot has been separated in two different parts for clarity. As explained in the previous chapter the number of stochastic COs are also drain voltage dependent: a high drain voltage is needed for COs to appear in a asymmetric double or multiple dots system [12,13]. It is obvious on Fig. 8, that no CO can be seen at $VD_{NW}$=4.5V, whereas more of them are evident at $VD_{NW}$=6V. The two drain voltages in-between showing intermediate behaviour.

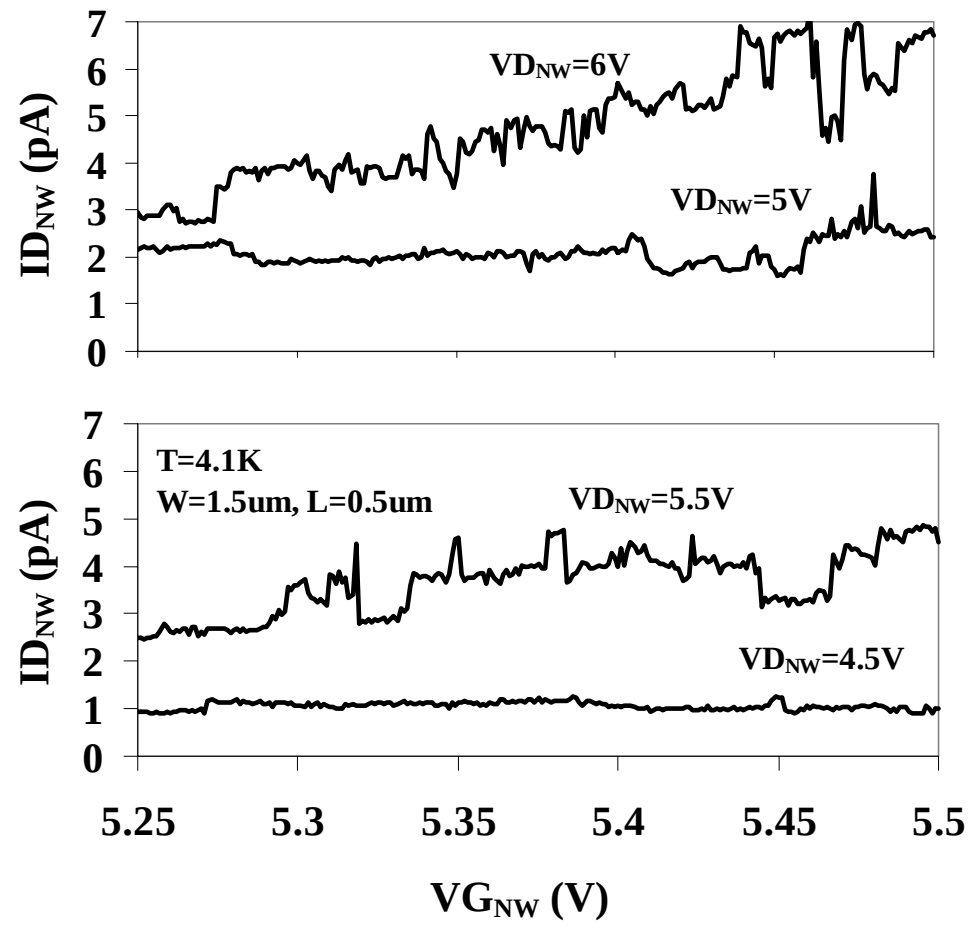

***Fig. 8:*** *measured $ID_{NW}$-$VG_{NW}$ characteristic of a polySiNW at 4.1K for various drain voltages (meas. step 1mV). The plot has been separated in two parts for clarity.*

Considering Fig. 6 to 8, it appears that Coulomb Blockade is effective in the gated polySiNW at low but intermediate temperature what is revealed by stochastic COs due to the randomly distributed grain sizes. Moreover those stochastic COs are drain voltage and temperature dependent which confirms previously reported results and MC simulations presented before.

## 4. CONCLUSION

Investigations of electrical characteristics of 10nm-thin polySiNW with nanograins (5 to 20nm) have been reported from 4K up to 300K. More effective COs have been observed at higher drain voltages and, especially, at intermediate range of temperatures (between 25K and 150K). MC simulations performed on an array of conductive islands connected to each other by tunnel junctions allow the validation of the experimental observations and the proposition of a first order theory.

### Acknowledgements

This work has been partly funded by the European Commission under the frame of the Network of Excellence "SINANO" (Silicon-based Nanodevices, IST-506844).